\pdfoutput = 1
\documentclass[aps,reprint,amsmath,amssymb,prb,superscriptaddress,floatfix,longbibliography,pdftex]{revtex4-2}
\usepackage[pdftex]{graphicx}
\usepackage{dcolumn}
\usepackage{bm}
\usepackage{braket}
\usepackage{color}
\usepackage{mathcomp}
\usepackage{indentfirst}
\usepackage{titlesec}
\titlespacing*{\section}
{0pt}
{0.3em}
{0.2em}
\usepackage[colorlinks,citecolor=blue,linkcolor=blue,urlcolor=blue]{hyperref}

\begin{document}

\title{Three-dimensional spin susceptibility in Ba$_{0.75}$K$_{0.25}$Fe$_{2}$As$_{2}$: Out-of-plane modulation revealed by neutron spectroscopy and theoretical modeling}
\author{Naoki Murai}
\email{naoki.murai@j-parc.jp}
\affiliation{Materials and Life Science Division, J-PARC Center, Japan Atomic Energy Agency, Tokai, Ibaraki 319-1195, Japan \looseness=-1}
\author{Katsuhiro Suzuki}
\email{Ka.Suzuki@niihama.kosen-ac.jp}
\affiliation{Department of Mechanical Engineering, National Institute of Technology, Niihama College, Niihama 792-8580, Japan}
\author{Masamichi~Nakajima}
\affiliation{RIKEN Center for Emergent Matter Science (CEMS), Wako 351-0198, Japan}
\author{Maiko Kofu}
\affiliation{Institute for Solid State Physics, The University of Tokyo, Kashiwa 277-8581, Japan}
\affiliation{Materials and Life Science Division, J-PARC Center, Japan Atomic Energy Agency, Tokai, Ibaraki 319-1195, Japan \looseness=-1}
\author{Seiko~Ohira-Kawamura}
\affiliation{Materials and Life Science Division, J-PARC Center, Japan Atomic Energy Agency, Tokai, Ibaraki 319-1195, Japan \looseness=-1}
\author{Yasuhiro~Inamura}
\affiliation{Materials and Life Science Division, J-PARC Center, Japan Atomic Energy Agency, Tokai, Ibaraki 319-1195, Japan \looseness=-1}
\author{Ryoichi Kajimoto}                                                                           
\affiliation{Materials and Life Science Division, J-PARC Center, Japan Atomic Energy Agency, Tokai, Ibaraki 319-1195, Japan \looseness=-1}

\date{\today}
\begin{abstract}
We present a combined experimental and theoretical investigation of the spin dynamics in the iron-based superconductor Ba$_{0.75}$K$_{0.25}$Fe$_2$As$_2$. 
Time-of-flight inelastic neutron scattering measurements reveal the three-dimensional (3D) nature of the spin fluctuations, manifested as out-of-plane modulations of the low-energy magnetic intensity.  
As the energy increases, this 3D-like modulation gradually fades away, leading to a more two-dimensional (2D) profile---a clear signature of a 3D-to-2D crossover in the spin dynamics. 
By incorporating a realistic 3D electronic band structure derived from density functional theory (DFT), we reproduce the experimentally observed features of the spin susceptibility, 
including the pronounced out-of-plane modulation at low energies and its gradual evolution into a more 2D character at higher energies. 
The calculated susceptibility exhibits a peak at the experimental ordering wavevector $\mathbf{q}_{\mathrm{AFM}} = (0.5, 0.5, 1)$, demonstrating that the DFT-derived 3D model accurately captures the tendency toward out-of-plane antiferromagnetic (AFM) order. 
Notably, electronic states away from the Fermi level play a crucial role in shaping the susceptibility peak at $\mathbf{q}_{\mathrm{AFM}}$, highlighting the limitations of the Fermi surface nesting picture in explaining the out-of-plane AFM instability. 
The demonstrated agreement between experiment and theory serves as a benchmark for validating the DFT-derived model as a realistic description of the material-specific electronic structure.
\end{abstract}
\maketitle

\section{INTRODUCTION}
\indent The discovery of iron-based superconductors (FeSCs) in 2008~\cite{Kamihara2008} generated considerable interest within the condensed matter physics community. 
Since then, extensive research efforts have been devoted to elucidating the mechanisms behind high-$\it T_{\rm c}$ superconductivity~\cite{Ishida2009, Paglione2010, Hirschfeld2011, Chubukov2015, Hirschfeld2016, Fernandes2022}.
In most FeSCs, superconductivity emerges in close proximity to an antiferromagnetic (AFM) phase, 
underscoring the potential role of spin fluctuations in the pairing mechanism. 
The idea of unconventional pairing mediated by spin fluctuations has a long history  
dating back to the seminal works on heavy fermion materials~\cite{Miyake1986,Scalapino1986} and later on the cuprates~\cite{Monthoux1991}. 
The discovery of FeSCs revived this line of research, and within a few years, 
a consensus formed around the spin-fluctuation-mediated pairing scenario~\cite{Mazin2008, Kuroki2008, Kuroki2009, Graser2009, Graser2010, Ikeda2010PR81.054502, Suzuki2011JPSJ, Suzuki2011PRB}. 
Today, FeSCs are considered prototypical model systems---alongside heavy fermion materials and cuprates---in which magnetism plays a fundamental role in the pairing mechanism~\cite{Scalapino2012}.\\
\indent In FeSCs, the low-energy bands near the Fermi level arise primarily from Fe 3$\it d$ orbitals, 
forming multiple hole and electron Fermi surfaces. 
Constructing a multi-orbital model that includes all five Fe 3$\it d$ orbitals is therefore an essential first step in theoretical treatment. 
The increasing sophistication of electronic structure theory has greatly facilitated the derivation of material-specific model Hamiltonians from density functional theory (DFT)~\cite{Marzari1997,Souza2001,Marzari2012}. 
Applied extensively to FeSCs, these first-principles techniques have matured into a well-established theoretical framework, 
enabling accurate predictions of magnetic instability in the stripe-type AFM channel~\cite{Mazin2008, Kuroki2008, Kuroki2009, Graser2009, Graser2010, Ikeda2010PR81.054502, Ikeda2010PRB82.024508, Kaneshita2010, Park2011, Suzuki2011JPSJ, Suzuki2011PRB,Yin2014}, 
consistent with neutron scattering experiments~\cite{Dai2012, Tranquada2014, Dai2015, Inosov2016, Zhao2008_a, McQueeney2008, Christianson2008, Ewings2008, Ishikado2009, Christianson2009, Harriger2009, Diallo2009, Zhao2009, Lester2010, Inosov2010, Diallo2010, Pratt2010, Wang2010, Park2010, Ewings2011, Zhang2011, 
Castellan2011, Harriger2011, Liu2012, Tucker2012, Luo2012, Harriger2012, Iimura2013, Luo2013, Wang2013, Tucker2014, Zhang2014, Lu2014, Qureshi2014, Rahn2015, Song2015, Kim2015, Wang2016Nat.Mater., Li2016, Wang2016Nat.Commun., Ding2016,
Xie2018, Sapkota2018, Murai2018, Lu2018, Guo2019, Chen2019, WaBer2019, Shen2020}. \\
\indent Despite this success, one crucial aspect has been largely overlooked---the three-dimensional (3D) character of magnetism. 
This aspect is broadly relevant to FeSCs, whose parent compounds, such as BaFe$_2$As$_2$, tend to exhibit 3D long-range AFM order 
with a wavevector $\mathbf{q}_{\mathrm{AFM}} = (0.5, 0.5, 1)$~\cite{Dai2015, Huang2008, Zhao2008_b, Kaneko2008, Goldman2008}.
The proximity to 3D AFM order suggests that the paramagnetic spin susceptibility---an indicator of the incipient magnetic instability---should also exhibit a non-negligible momentum dependence, not only within the plane but also along the out-of-plane direction.
However, most theoretical studies to date have primarily focused on the in-plane momentum dependence of spin susceptibility, often assuming a quasi-two-dimensional (2D) electronic structure.
A notable exception is the pioneering work by Park \textit{et al.}~\cite{Park2010}, who examined out-of-plane momentum dependence of the spin susceptibility in BaFe$_2$As$_2$---one of the few studies to address the 3D aspect of magnetism. 
Nonetheless, their study leaves several aspects open to further scrutiny.
Experimentally, they only investigated the low-energy region, leaving the out-of-plane spin susceptibility at higher energies largely unexplored.
Theoretically, their analysis was restricted to the static limit of the irreducible susceptibility, which does not directly correspond to the dynamical spin susceptibility measured in inelastic neutron scattering (INS) experiments. 
Moreover, their irreducible susceptibility along the out-of-plane \((0.5, 0.5, L)\) direction exhibits a peak slightly offset from $L$ = 1 for BaFe$_{2}$As$_{2}$, 
failing to capture the experimentally observed maximum associated with the out-of-plane AFM ordering.
This limitation of previous approaches highlights the need to thoroughly revisit the 3D dynamical spin susceptibility, 
directly comparable to INS results.
\\
\indent Here, we bridge this gap by combining time-of-flight (TOF) neutron spectroscopy with DFT-based modeling of the spin susceptibility, 
which captures the full 3D momentum dependence, including its out-of-plane variation.
A central question we address is whether DFT-derived models can reliably reproduce the experimentally observed momentum dependence of the spin susceptibility---both in-plane and out-of-plane---thereby 
offering a stringent test of these models via comparison with experiment. 
INS measurements reveal a clear modulation of the low-energy magnetic signal along the \((0.5, 0.5, L)\) direction, with peaks at odd $L$ positions 
reflecting out-of-plane AFM instability.  
As the energy increases, this 3D-like intensity modulation gradually fades away, resulting in a nearly 2D magnetic response. 
Using a realistic 3D electronic structure model derived from DFT calculations, we reproduce the $L$-modulation of the low-energy spin susceptibility, consistent with the observed out-of-plane AFM correlations.
Moreover, our calculations capture its gradual suppression at higher energies, revealing a 3D-to-2D crossover in the spin susceptibility. 
Together, these findings provide a stringent test of DFT-derived 3D models and affirm their validity in describing the momentum-dependent spin dynamics of FeSCs.
\\
\section{EXPERIMENTAL METHODS}
\indent Single crystals of Ba$_{0.75}$K$_{0.25}$Fe$_{2}$As$_{2}$ were grown using the FeAs-flux method~\cite{Nakajima2018, Murai2018}. 
The room-temperature $c$-axis lattice parameter was determined to be 13.212 {\AA{}}, 
indicating a potassium concentration of approximately 25{\%}, consistent with Refs.~\cite{Rotter2008, Kihou2016}.
For neutron scattering measurements, the crystals were co-aligned on several aluminum plates using a hydrogen-free adhesive (CYTOP), resulting in an array with a total mass of 5.0~g. 
The N\'eel and superconducting transition temperatures, determined from neutron diffraction measurements on the co-aligned array, are $T_{\rm N}$ = 90~K and $T_{\rm c}$ = 25~K, respectively (see Supplemental Material for 
details of the neutron diffraction characterization; see also Refs.~\cite{Waber2015,Avci2014,Bohmer2015,Taddei2016,Allred2016} therein).\\
\indent INS measurements were carried out on the co-aligned array of Ba$_{0.75}$K$_{0.25}$Fe$_2$As$_2$ single crystals to obtain the dynamic structure factor 
$S(\mathbf{Q}, \omega)$, which encodes momentum- and energy-resolved information about spin dynamics. 
These measurements were conducted using the AMATERAS~\cite{Nakajima2011} and 4SEASONS~\cite{Kajimoto2011, Nakamura2009} TOF chopper spectrometers at the 
Materials and Life Science Experimental Facility of J-PARC.
For the AMATERAS measurements, we used incident neutron energies ($E_i$) of 7.74 (0.25), 15.15 (0.56), and 42.0 (2.4)~meV, 
where the numbers in parentheses denote the energy resolution at the elastic line. 
The co-aligned crystal array was mounted with the tetragonal $(H, H, L)$ scattering plane oriented horizontally. 
To reconstruct the four-dimensional (4D) scattering function $S(\mathbf{Q}, \omega)$, a series of data sets was collected at multiple orientations 
by rotating the crystal array in 1$^\circ$ steps over the range $\varphi \in [-40^\circ, +40^\circ]$, 
where $\varphi$ denotes the angle between the incident neutron beam and the $c$-axis of the crystal. 
In addition to this discrete step-scan method, we employed a recently developed continuous rotation scan technique, 
wherein the 4D $S(\mathbf{Q}, \omega)$ was reconstructed from a single data set acquired during uninterrupted rotation. 
For the 4SEASONS measurements, we used higher incident neutron energies of 55.6 (2.9) and 125 (9)~meV to extend the accessible region in 4D $(\mathbf{Q},\omega)$ space.
The same crystal array was again mounted in the $(H, H, L)$ geometry and rotated over the range $\varphi \in [-20^\circ, +60^\circ]$ in 0.5$^\circ$ steps, 
with data collected separately at each angle.
Additionally, fixed-orientation measurements at $\varphi = 0^\circ$ were conducted with $E_i = 31.3$ (1.5)~meV in the $(H, 0, L)$ geometry, 
under the assumption of negligible $L$-dependence.
We used the \textsc{Utsusemi} and \textsc{D4mat2} software packages~\cite{Inamura2013, Utsusemi} 
to convert the above-obtained neutron event data into the momentum- and energy-resolved scattering function $S(\mathbf{Q}, \omega)$.
The momentum transfer $\mathbf{Q}$ is expressed in reciprocal lattice units (r.l.u.)\ of the tetragonal unit cell, with 
\({\mathbf{Q}} = H\mathbf{a}^{*} + K\mathbf{b}^{*} + L\mathbf{c}^{*} \equiv (H, K, L)\). 
In this notation, low-energy spin fluctuations are typically found near $\mathbf{Q} = (0.5, 0.5, L)$, and 
we focus on their structure along the out-of-plane $L$ direction.
\\
\section{THEORETICAL METHODS}
\indent Given the metallic nature of FeSCs, an itinerant approach to modeling spin fluctuations is an appropriate choice~\cite{Ikeda2010PRB82.024508,Luo2012,Lu2018,Murai2018}. 
In this work, we evaluate the dynamical spin susceptibility within the multiorbital random phase approximation (RPA) formalism using a realistic DFT-derived band structure.\\
\indent The unit cell of BaFe$_{2}$As$_{2}$ contains two Fe atoms due to the presence of two inequivalent As positions, requiring a ten-orbital model for an accurate description of the electronic band structure. 
However, since the magnetic moment is localized on the Fe atoms, 
the INS signals reflect the symmetry of the unfolded Brillouin zone for the Fe sublattice (1-Fe/unit cell), 
rather than that of the crystallographic unit cell (2-Fe/unit cell)~\cite{Park2010,Park2011}. 
Consequently, a five-orbital model in the 1-Fe/unit cell Brillouin zone provides a more appropriate starting point for describing spin susceptibility in the paramagnetic phase.
With this in mind, we constructed an effective five-orbital tight-binding model of BaFe$_{2}$As$_{2}$ by unfolding the ten-orbital model into an effective 1-Fe/unit cell Brillouin zone~\cite{Suzuki2011PRB}. 
To obtain the tight-binding model from first principles, we used the {\scshape Quantum ESPRESSO}~\cite{Giannozzi2009,Giannozzi2017} and {\scshape Wannier90}~\cite{Mostofi2008,Pizzi2020} software packages. 
The DFT calculations were performed using the generalized gradient approximation (GGA) exchange-correlation functional~\cite{Perdew1996} 
with a cutoff energy of 40 Ry and a $k$-point mesh of $8\times 8\times 8$.
The K substitution effect in Ba$_{0.75}$K$_{0.25}$Fe$_{2}$As$_{2}$ was modeled as a rigid band shift of the Fermi level. 
To account for the experimentally observed band narrowing due to electron correlations, 
we rescaled the DFT band structure by a factor of three to match the angle-resolved photoemission spectroscopy (ARPES) data~\cite{Murai2018,Liu2008}. 
The dynamical spin susceptibility was then obtained using the RPA: 
\begin{align}
  \label{eq1}
  \hat{\chi}_s(\mathbf{q},\omega)=\hat{\chi}_0(\mathbf{q},\omega)[\hat{I}-\hat{S}\hat{\chi}_0(\mathbf{q},\omega)]^{-1}, 
  \end{align}
 where \(\hat{S}\) is the interaction vertex matrix~\cite{Yada2005}. 
 The irreducible susceptibility \(\hat{\chi}_0(\mathbf{q},\omega)\) is given as  
 \begin{align}
  \label{eq2}
  \hat{\chi}^{l_1,l_2,l_3,l_4}_0&(\mathbf{q},\omega)=\sum_k\sum_{n,m}\frac{f(\varepsilon^n_{\mathbf{k}+\mathbf{q}})-f(\varepsilon^m_{\mathbf{k}})}{\omega-\varepsilon^n_{\mathbf{k+q}}+\varepsilon^m_{\mathbf{k}}+i\delta}\\\nonumber
  &\times U_{l_1,n}(\mathbf{k+q})U_{l_4,m}(\bm{k})U^\dagger_{m,l_2}(\bm{k})U^\dagger_{n,l_3}(\mathbf{k+q}),
\end{align}
where $f$, $\varepsilon^m_{\mathbf{k}}$, and $U_{l,m}(\mathbf{k})$ represent the Fermi distribution function, 
the eigenvalue of the Bloch state with momentum $\mathbf{k}$ and band index $m$, 
and the matrix element of the unitary transformation that connects the orbital and band spaces, respectively.
The orbital index \(l \in (1,...,5)\) corresponds to the Fe 3$\it d$ orbitals \((d_{xy}, d_{xz}, d_{yz}, d_{x^2-y^2}, d_{z^2})\). 
As the elements of $\hat{S}$, we consider the Hubbard-type interactions, i.e., intra- and inter-orbital onsite interaction $U$, $U'$, Hund's coupling $J$, and pair hopping $J'$. 
The calculation was performed on a $k$-point mesh of $128\times 128\times 128$ with $U =0.41$ eV, $U'=U-2J$, $J=J'=U/8$, 
temperature of $k_{\rm B}T=3.0\times 10^{-2}$~eV, and smearing factor of $\delta=5.0\times 10^{-3}$~eV.
\\
\section{RESULTS AND DISCUSSION}
\begin{figure}[t]
  \begin{center}
    \includegraphics[width=\linewidth, pagebox=artbox]{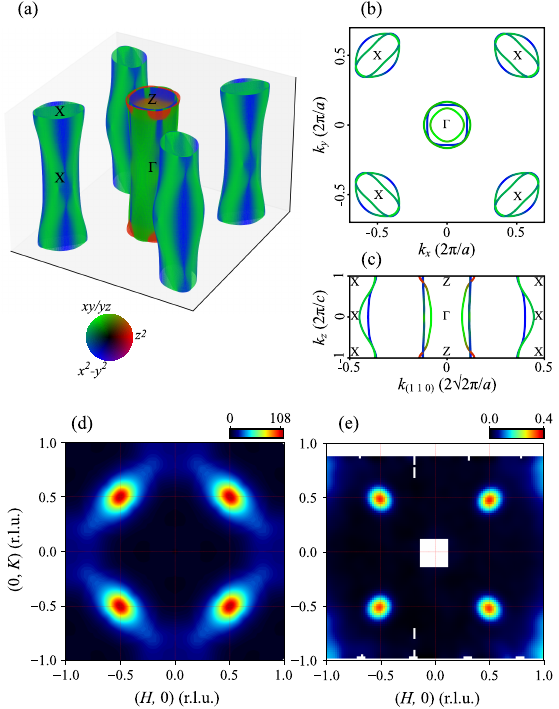}
    \caption{(a)--(c) 3D and cross-sectional views of the DFT-derived Fermi surface for BaFe$_2$As$_2$, with the Fermi level shifted to account for $25\%$ K-doping. 
      Panel (a) shows a 3D view, (b) a top view from the (0 0 1) direction, and (c) a side view from the (1 $\bar 1$ 0) direction. 
      The color code represents the orbital character projected onto the Fermi surface. 
      Panels (d) and (e) show the theoretical and experimental spin susceptibilities at $\omega = 10$~meV, respectively, both peaking at \(\mathbf{q} = (0.5, 0.5)\). 
      The data in panel (e) were obtained from a fixed-geometry scan ($\varphi = 0^\circ$), where the sample $c$-axis was aligned parallel to the incident neutron beam. 
      The measurements were performed using $E_i = 31.3$~meV at $T = 30$ K.
      The intensities in panels (d) and (e) are shown on different scales.
    }
  \label{Fig1}
  \end{center}
\end{figure}

\indent To set the stage, we briefly review the magnetic instability inherent in the Fermi surface geometry of FeSCs. 
Figures~\ref{Fig1}(a)-(c) show the orbital-resolved Fermi surface for $25\%$ K-doped BaFe$_{2}$As$_{2}$~\footnote{
The doping effect is modeled through a rigid band shift of the Fermi level. 
To follow the convention often used in prior studies, the Fermi surfaces in Figs.~\ref{Fig1}(a)-(c) are shown in the folded Brillouin zone for the crystallographic unit cell (2-Fe/unit cell).
For the RPA calculations, however, we used an effective five-orbital model in the unfolded Brillouin zone of the 1-Fe/unit cell, which better reflects the symmetry of spin susceptibility in FeSCs.
}, featuring hole pockets centered at the Brillouin zone center ($\mathrm{\Gamma}$) and electron pockets at the zone corner ($\mathrm{X}$). 
The well-nested nature of these pockets implies an enhanced low-energy spin susceptibility at the nesting wavevector $\mathbf{q} = (0.5, 0.5)$, 
which connects $\Gamma$ and $X$. 
Explicit calculations based on the DFT-derived model indeed reveal a pronounced susceptibility maximum at this wavevector [Fig.~\ref{Fig1}(d)], 
consistent with the experimentally observed peak position in the INS data [Fig.~\ref{Fig1}(e)]. 
(For a detailed discussion of the peak shapes, see Ref.~\footnote{
The DFT-derived Fermi surface reproduces the disconnected topology of the hole and electron pockets, 
although minor quantitative differences remain in their relative sizes and detailed geometry. 
The susceptibility peak at $\mathbf{q}=(0.5,0.5)$ arises from scattering between these disconnected sheets. 
Because this Fermi surface topology is robustly reproduced within DFT, the corresponding peak position (i.e., the instability wavevector) is captured at the correct momentum. 
By contrast, the detailed peak shape reflects finer geometrical features of the Fermi surface---most notably the subtle size mismatch between the hole and electron pockets---and is 
therefore described only qualitatively within the present approximation 
(see Sec.~III of the Supplemental Material for a rigid-band analysis of the doping-induced Fermi surface size mismatch and its impact on the peak anisotropy; see also Refs.~\cite{Miyake2010} therein.).
}.)
\\
\begin{figure*}[t]
  \begin{center}
    \includegraphics[width=0.98\linewidth, pagebox=artbox]{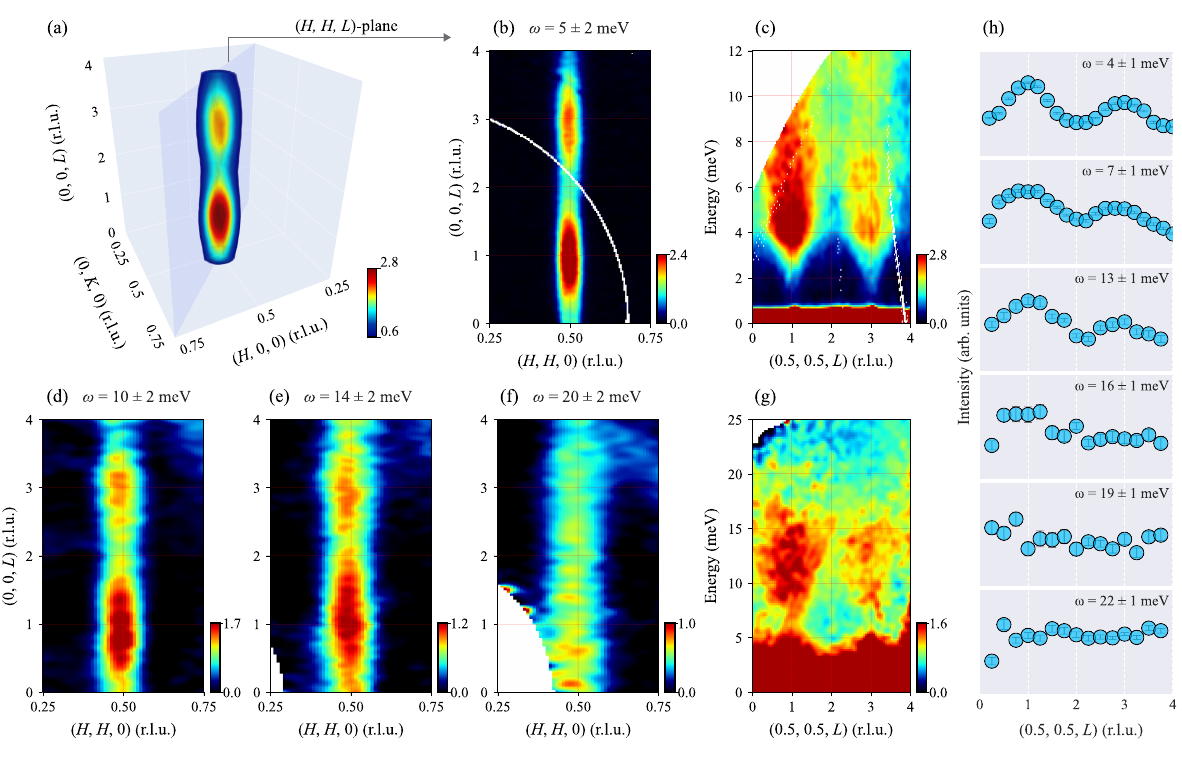}
    \caption{
  (a--c) Low-energy spin excitations measured with $E_i = 15.15$~meV, focusing on the out-of-plane ($L$) dependence of the magnetic response.
(a) 3D density map of the magnetic scattering intensity in $(H,K,L)$ space at $\omega = 5 \pm 2$ meV.
(b) Horizontal slice of panel (a) in the $(H, H, L)$ plane.
(c) False-color energy spectrum along $(0.5, 0.5, L)$.
Panels (a)--(c) reveal a pronounced periodic intensity modulation along $L$, with maxima at odd $L$ and minima at even $L$, thereby establishing the 3D character of the low-energy spin excitations.
(d--g) Spin excitations measured with a higher incident energy of $E_i = 55.6$~meV.
(d--f) Constant-energy slices in the $(H, H, L)$ plane at $\omega = 10 \pm 2$, $14 \pm 2$, and $20 \pm 2$~meV.
While the odd-$L$ modulation remains clearly visible at 10 and 14 meV, it becomes strongly suppressed at 20 meV, where the intensity approaches a nearly uniform distribution along $L$.
(g) False-color energy spectrum along $(0.5,0.5,L)$, showing that the $L$-dependent modulation progressively weakens above $\omega \sim 15\,\mathrm{meV}$.
(h) Representative constant-energy cuts along $(0.5,0.5,L)$ at $\omega = 4 \pm 1$, $7 \pm 1$ meV ($E_i = 15.15$~meV) and $\omega = 13 \pm 1$, $16 \pm 1$, $19 \pm 1$, and $22 \pm 1$~meV ($E_i = 55.6$~meV), 
highlighting the crossover from a strongly $L$-modulated 3D response at low energies to a nearly $L$-independent 2D profile at higher energies.
All data were collected at $T < 10$ K. 
}
  \label{Fig2}
\end{center}
\end{figure*}
\indent
The DFT-derived model thus captures the in-plane susceptibility peak at $\mathbf{q} = (0.5,0.5)$, characteristic of the stripe-type AFM instability. 
To fully understand the magnetic response, however, one must also resolve its out-of-plane momentum dependence.
As shown in Fig.~\ref{Fig1}(c), the Fermi surface exhibits noticeable $k_z$ warping and variations in orbital character, 
indicating a subtle yet non-negligible three-dimensionality in the electronic structure (see also ARPES confirmation~\cite{Vilmercati2009,Malaeb2009,Yoshida2011,Brouet2012,Ideta2013,Ye2013,Suzuki2014,Ideta2019}). 
Given the sensitivity of spin susceptibility to this three-dimensionality, it is essential to examine its behavior along $\mathbf{q} = (0.5, 0.5, L)$. 
Nevertheless, most studies have focused on magnetic excitations within the 2D $(H, K)$ plane, leaving the out-of-plane response largely unexplored.
This limited focus stems from the prevailing assumption that the spin susceptibility of FeSCs is largely independent of $L$, 
reflecting their quasi-2D electronic structure. 
Consequently, earlier TOF INS studies on FeSCs often employed fixed-geometry scans suitable for 2D materials.  
In this approach, \(S(\mathbf{Q}, \omega)\) is obtained by projecting the observed signal onto the \((H, K, E)\) coordinate system 
while assuming implicit $L$-independence---an assumption that risks overlooking out-of-plane spin dynamics arising from the 3D electronic structure of the material.
\\
\begin{figure*}[t]
  \begin{center}
    \includegraphics[width=0.97\linewidth, pagebox=artbox]{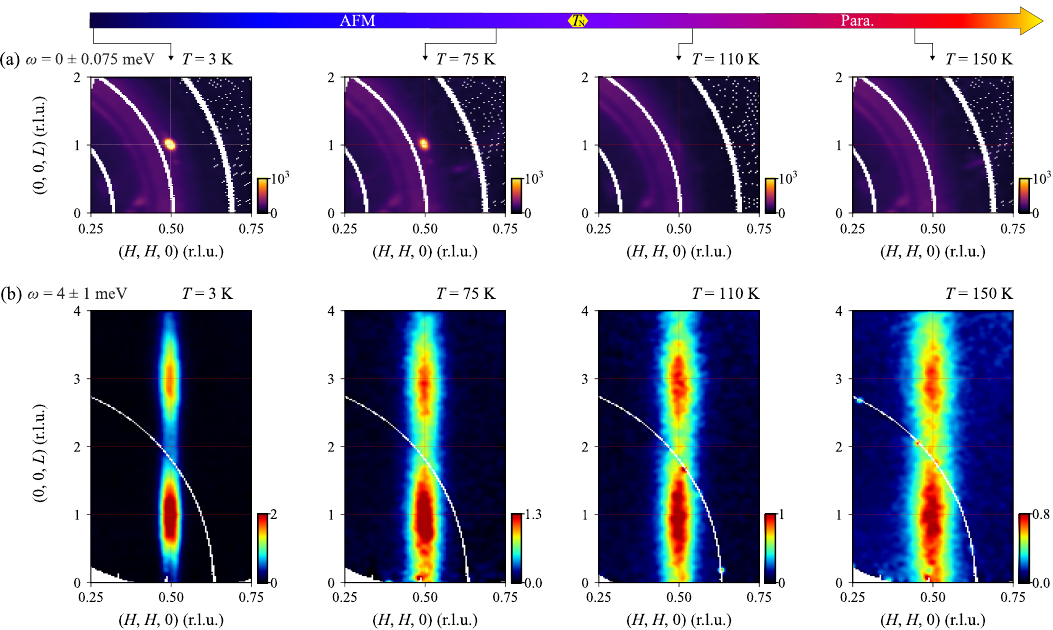}
    \caption{
Temperature dependence of magnetic scattering in the \((H, H, L)\) plane at (a) $\omega = 0 \, \mathbin{\pm} \, 0.075$~meV and (b) $\omega = 4 \, \mathbin{\pm} \, 1$~meV, respectively, 
collected with $E_i = 7.74$~meV. 
At finite energy transfer in panel (b), magnetic scattering peaks appear at odd $L$ positions and persist above $T_\mathrm{N}$, 
indicating robust out-of-plane AFM correlations.
}
  \label{Fig3}
\end{center}
\end{figure*}
\indent To overcome this limitation, we performed measurements of the full 4D $S(\mathbf{Q}, \omega)$ by combining large data sets collected from multiple crystal orientations. 
Unlike fixed-geometry scans, which cover only a sparse set of hypersurfaces in the 4D $(\mathbf{Q}, \omega)$ space, 
this multi-orientation approach substantially expands the accessible region, thereby enabling comprehensive measurement of the full 4D volume~\cite{Fernandez2013, Boothroyd2020}. 
As a demonstration, we generated a 3D density map of the magnetic signal in the $(H, K, L)$ space at $\omega = 5 \pm 2$ meV [Fig.~\ref{Fig2}(a)]. 
The magnetic signal centered at $(0.5, 0.5)$ extends in a rodlike pattern along the $L$ direction, as expected for a quasi-2D system. 
However, the intensity distribution varies significantly with $L$. 
To better visualize these variations, we sliced the 3D density map along the 2D $(H, H, L)$ plane [Fig.~\ref{Fig2}(b)]. 
The magnetic signal is enhanced at odd $L$ positions and suppressed at even $L$ positions, forming a periodic intensity modulation along the $\it L$ direction. 
The corresponding energy spectrum along $(0.5,0.5,L)$ [Fig.~\ref{Fig2}(c)] exhibits the same odd-$L$ enhancement, confirming the presence of the $L$-dependent modulation in the energy spectrum.
Similar modulations have also been reported in other 122 systems, 
including BaFe$_{2}$As$_{2}$~\cite{Harriger2009, Pratt2010, Park2010, Zhang2011, Shen2020}, 
SrFe$_{2}$As$_{2}$~\cite{Zhao2008_a, Guo2019} and 
CaFe$_{2}$As$_{2}$~\cite{McQueeney2008, Sapkota2018}. 
However, such investigations were limited to low-energy excitations, and the behavior at higher energies remained largely unexplored.
To investigate how this $L$-dependent modulation evolves with increasing energy, we performed additional measurements with $E_i = 55.6$~meV.
Constant-energy slices in the $(H,H,L)$ plane at $\omega = 10 \pm 2$, $14 \pm 2$, and $20 \pm 2$~meV are shown in Fig.~\ref{Fig2}(d)--(f).
The odd-$L$ modulation, clearly visible at 10 and 14~meV, is strongly suppressed at 20~meV, with the intensity approaching a nearly uniform distribution along $L$.
This evolution is more clearly revealed in the energy spectrum along $(0.5,0.5,L)$ [Fig.~\ref{Fig2}(g)], which shows that the pronounced 3D modulation observed at low energies weakens continuously above $\omega \sim 15\,\mathrm{meV}$.
Representative constant-energy cuts [Fig.~\ref{Fig2}(h)] further demonstrate that the distinct odd-$L$ peaks gradually collapse into a nearly $L$-independent profile at higher energies
\footnote{ Additional measurements with a higher incident energy of $E_i = 125 $ meV extend the accessible energy range beyond that covered by $E_i = 55.6$ meV. 
These data likewise exhibit a nearly $L$-independent intensity distribution at high energies, further corroborating the 2D character of the out-of-plane spin fluctuations (see Supplemental Material).
}.
Taken together, these results establish a continuous crossover from a strongly $L$-modulated 3D response at low energies to a nearly 2D response at higher energies.
\\
\indent We now focus on the low-energy regime, where the 3D nature of spin fluctuations manifests most prominently. 
Figures~\ref{Fig3}(a) and \ref{Fig3}(b) present the temperature dependence of the magnetic scattering in the \((H, H, L)\) plane. 
As shown in Fig.~\ref{Fig3}(a), the magnetic Bragg peak at the odd $\it L$ position disappears once the temperature exceeds $\it T_{\rm N}$.  
At finite energy transfer $\omega = 4 \pm 1$ meV [Fig.~\ref{Fig3}(b)], the magnetic signal also peaks at odd $L$ positions---much like the magnetic Bragg peak. 
However, its $L$-modulated momentum dependence persists even above $T_{\mathrm{N}}$.
A similar $L$-dependent magnetic signal has been reported to persist against doping-induced suppression of the AFM order~\cite{Park2010, Zhang2011, Shen2020}. 
Thus, regardless of whether the AFM order is suppressed by temperature or doping, the out-of-plane magnetic signal remains peaked at odd $\it L$ positions, demonstrating the robustness of out-of-plane AFM correlations.
Such behavior is consistent with the expectation that the paramagnetic spin susceptibility---a direct measure of the magnetic ordering tendency---should peak at the AFM wavevector \(\mathbf{q}_{\rm AFM} = (0.5, 0.5, 1)\), 
rather than exhibit a uniform intensity profile along $L$, as would be expected for a purely 2D system.
Using the first-principles model, we successfully reproduce the in-plane spin susceptibility peaking at \(\mathbf{q} = (0.5, 0.5)\) [see Fig. \ref{Fig1}(d)]. 
The key question here is whether our theoretical approach can capture the observed $L$-modulation reflecting the out-of-plane AFM instability, together with its energy-dependent suppression that leads to the 3D-to-2D crossover.
\begin{figure}[t]
  \begin{center}
    \includegraphics[width=\linewidth, pagebox=artbox]{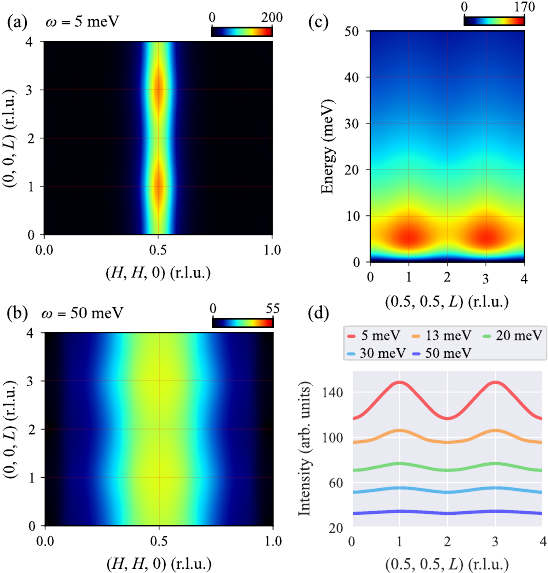}
    \caption{
      (a) Constant-energy map of the RPA spin susceptibility at $\omega = 5$ meV in the $(H, H, L)$ plane, calculated using a DFT-derived 3D band structure. 
      The intensity exhibits a periodic modulation along $L$, with maxima at odd $L$ and minima at even $L$ positions.
      (b) Corresponding map at $\omega = 50$ meV, showing a nearly uniform intensity distribution along $L$.
      (c) Energy dependence of the RPA spin susceptibility along the $(0.5, 0.5, L)$ direction.
      (d) Constant-energy cuts along $(0.5, 0.5, L)$ at $\omega = 5$, 13, 20, 30, and 50 meV, highlighting the strong $L$-dependent intensity variation at low energies and its gradual suppression at higher energies.
      }
  \label{Fig4}
  \end{center}
  \end{figure}
\\
\indent Motivated by this, we computed the out-of-plane spin susceptibility using RPA.  
To fully account for the explicit $L$ dependence, we considered the full 3D momentum dependence of the electronic band structure.  
Figure~\ref{Fig4}(a) shows a constant-energy map at \(\omega = 5\) meV in the \((H, H, L)\) plane.  
The spin susceptibility exhibits a periodic intensity modulation along the $L$ direction, with maxima at odd $L$ and minima at even $L$ positions.  
At a higher energy of \(\omega = 50\) meV, the spin susceptibility exhibits a nearly 2D intensity distribution, with the $L$ modulation significantly suppressed, as shown in Fig.~\ref{Fig4}(b).  
The energy spectrum in Fig.~\ref{Fig4}(c) further confirms the presence of the $L$-dependent intensity modulation, 
which is most pronounced in the low-energy regime and gradually suppressed at higher energies. 
Its energy evolution is more clearly visualized in Fig.~\ref{Fig4}(d), where constant-energy cuts along the $(0.5, 0.5, L)$ direction 
reveal a gradual crossover from pronounced $L$ modulation at low energies to an almost uniform intensity distribution at higher energies. 
At \(\omega = 5\) meV, the intensity at $L$ = 1 is about 28\% stronger than that at $L$ = 0, whereas the difference is only 6\% at \(\omega = 50\) meV, 
thereby capturing the crossover from 3D to nearly 2D spin susceptibility, in agreement with the INS results.\\
\indent Thus, our approach successfully reproduces the key features of the observed $L$-modulated spin susceptibility over a wide energy range, 
offering a more complete description than the previous study~\cite{Park2010}, which only considered the static irreducible susceptibility.
As described in Eqs.~(\ref{eq1})--(\ref{eq2}), the dynamical spin susceptibility \(\hat{\chi}_s(\mathbf{q},\omega)\) is obtained by summing over electronic states, making its momentum dependence highly sensitive to the underlying band structure. 
In an ideal 2D system with no \(k_z\) dispersion, integrating over \(k_z\) does not introduce any $\it L$ dependence, resulting in a uniform intensity.
In contrast, when the electronic structure exhibits explicit \(k_z\) dispersion, the \(k_z\)-integrated contributions to \(\hat{\chi}_s(\mathbf{q},\omega)\) lead to an $L$-dependent intensity variation. 
Thus, a 3D modulation of the spin susceptibility follows directly from the underlying 3D electronic band structure.
The agreement with the experiment supports the validity of our DFT-derived model as a reliable description of the 3D electronic structure.\\
\indent One might be tempted to attribute the observed $L$ modulation of the spin susceptibility---manifest only at low energies---to a much weaker interlayer exchange interaction than the in-plane couplings, 
as assumed in a localized spin picture. 
Such strong-coupling descriptions can reproduce some magnetic features of FeSCs, but an itinerant approach---one that directly incorporates the 
band structure and Fermi surface geometry---is better suited to capture the magnetism and its interplay with superconductivity~\cite{Hirschfeld2011, Hirschfeld2016}. 
Accordingly, we interpret the out-of-plane spin dynamics within this framework to maintain theoretical consistency.
Our itinerant approach yields an \textit{a priori} prediction of the 3D-to-2D crossover in spin susceptibility based on a material-specific DFT model, 
thus avoiding the \textit{ad hoc} exchange-path choices and \textit{post hoc} coupling tuning, as often required in localized spin models.
\\
\begin{figure}[t]
  \begin{center}
    \includegraphics[width=\linewidth, pagebox=artbox]{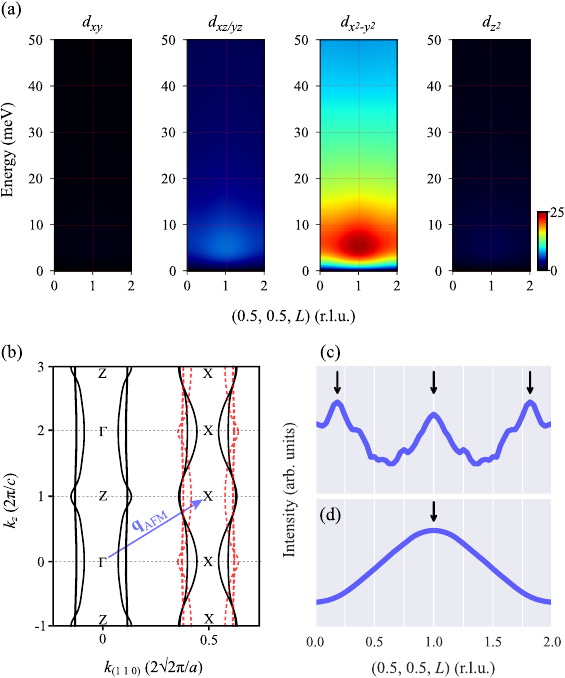}
    \caption{
      (a) Orbital-resolved intraorbital components of the RPA spin susceptibility, showing a common momentum dependence with a peak at $L = 1$ across all orbitals.
      (b) Schematic illustration of the imperfect nesting between hole and electron pockets. Red dotted lines indicate the hole Fermi surfaces translated by the AFM wavevector $\mathbf{q}_{\mathrm{AFM}} = (0.5, 0.5, 1)$ (blue arrow).
      (c) The spin susceptibility along the \((0.5, 0.5, L)\) direction, computed using only electronic states within \(\pm5\) meV of the Fermi level, thereby isolating the contribution from the Fermi surface geometry.
      (d) The spin susceptibility at \(\omega = 5\) meV, obtained using the full DFT band structure without any energy cutoff, thereby including contributions from a broad energy range. 
      Black arrows indicate the peak positions along $L$: 0.19, 1, and 1.81 in (c), and 1 in (d).
      }
  \label{Fig5}
  \end{center}
  \end{figure}
\indent To gain further insight into the origin of the $L$ modulation, 
we analyzed the orbital-resolved components of the spin susceptibility. 
As shown in Fig.~\ref{Fig5}(a), the overall intensity varies among orbitals, with the $d_{x^2-y^2}$ and $d_{xz/yz}$ components exhibiting the strongest signals, 
reflecting their predominant weight near the Fermi level.
Importantly, all components exhibit a common momentum dependence, with a peak at $L = 1$. 
This observation suggests that the modulation originates not from any particular orbital character, but rather from the overall 3D band structure. 
In this context, a first interpretation would attribute the peak at $L=1$ to nesting between 3D-warped Fermi surfaces along $k_z$.
However, as shown in Fig.~\ref{Fig5}(b), the Fermi surface geometry does not support this nesting scenario: translating the hole pockets by the AFM wavevector $\mathbf{q}_\mathrm{AFM} = (0.5, 0.5, 1)$ 
fails to produce substantial geometric overlap with the electron pockets, thereby challenging any attempt to account for the observed susceptibility peak solely from the Fermi surface geometry.
Similar inconsistencies have also been noted in an early ARPES study~\cite{Yoshida2011}. 
To further assess the role of Fermi surface geometry, we computed the spin susceptibility by evaluating Eqs.~(\ref{eq1})--(\ref{eq2}) using only electronic states within \(\pm 5\) meV of the Fermi level. 
This energy window restricts the contributions to low-energy electronic states, thereby isolating the influence of the Fermi surface geometry on the spin susceptibility. 
In this case, multiple other peaks appear at $L = 0.19$ and $1.81$, in addition to a peak at $L = 1$ [see Fig.~\ref{Fig5}(c)]. 
In contrast, when the full energy range of the band structure is taken into account, a single peak emerges at $L = 1$ [see Fig.~\ref{Fig5}(d)], thereby uniquely selecting the dominant magnetic instability at \(\mathbf{q}_{\rm AFM}\), consistent with the experimental observations. 
This comparison demonstrates that a substantial portion of the spin susceptibility originates from electronic states away from the Fermi level, and that 
a simple nesting picture, which considers only the Fermi surface geometry, is insufficient to explain the tendency toward out-of-plane AFM order. 
Visual inspection of the Fermi surface alone can be misleading. 
Accordingly, a reliable identification of the dominant magnetic instability requires explicit calculations of the spin susceptibility based on the full 3D band structure.
\\
\indent In closing, we offer some remarks on the implications of our results and remaining challenges for future studies of spin fluctuations in FeSCs.
The superconducting gap symmetry and the formation of spin fluctuations in FeSCs are intimately tied to the underlying electronic band structure. 
Hence, establishing the reliability of theoretical models is necessary, particularly for elucidating the pairing mechanism.
As demonstrated in this study, RPA combined with DFT-derived band structures successfully describes the out-of-plane spin susceptibility in 122 compounds, 
providing a reliable 3D electronic structure model applicable to these systems. 
Such a validated 3D model is particularly important for theoretical investigations of superconductivity in 122 compounds, where the $k_{\mathrm z}$-dependent 
electronic structure gives rise to strongly 3D gap features such as horizontal nodes~\cite{Graser2010,Suzuki2011JPSJ,Suzuki2011PRB}.
However, the success of our approach does not guarantee its applicability to more strongly correlated FeSCs, 
where electron correlations beyond the LDA/GGA level may play a crucial role. 
In this context, more advanced methods such as dynamical mean field theory (DMFT)---which has proven effective in reproducing in-plane spin excitations~\cite{Park2011, Yin2014}---may offer improved accuracy for describing out-of-plane spin correlations as well, although such extensions remain to be explored. 
Given its pronounced electronic correlations, FeSe stands out as a compelling case for testing beyond-DFT approaches~\cite{Aichhorn2010,Watson2017,Mandal2017,Acharya2022}. 
In this regard, we note a recent report of out-of-plane ferromagnetic (FM) spin fluctuations in FeSe~\cite{Ma2024}, which, to date, have been observed only in FeSe among FeSCs. 
This observation raises a key question: can standard DFT-derived models capture the observed out-of-plane FM instability, 
or must we turn to beyond-DFT approaches? 
Systematic validation of theoretical models against experiment remains a key task in the study of unconventional superconductors.
Such efforts will enhance the predictive power of theoretical frameworks and advance our understanding of unconventional pairing mechanisms in FeSCs. 
\\
\section{SUMMARY}
\indent In this study, we investigated the 3D character of spin fluctuations in Ba$_{0.75}$K$_{0.25}$Fe$_{2}$As$_{2}$ using a combined experimental and theoretical approach.
TOF neutron spectroscopy revealed an $L$-dependent modulation of the low-energy magnetic scattering, 
with intensity maxima at odd $L$ positions reflecting out-of-plane AFM correlations. 
As the energy increases, this 3D feature gradually fades away, signaling a 3D-to-2D crossover in the spin dynamics.
Constructing a realistic 3D model based on DFT calculations, we reproduce the key features of the observed spin susceptibility, 
including out-of-plane intensity modulation and its energy-dependent suppression.
Further analysis demonstrates that the observed $L$-modulation cannot be explained solely by orbital-selective effects or simple Fermi surface nesting. 
Instead, electronic states away from the Fermi level play a key role in selecting the susceptibility peak at $L = 1$, 
consistent with the experimentally observed out-of-plane AFM correlations. \\
\indent Capturing the 3D magnetic ordering tendency at \(\mathbf{q}_{\rm AFM} = (0.5, 0.5, 1)\)---a feature 
inaccessible to purely 2D models---requires a fully 3D treatment of the electronic structure. 
Using the DFT-derived 3D band structure, our itinerant approach reproduces the 3D-to-2D crossover without phenomenological parameter tuning, 
thus requiring neither downfolding to an effective spin Hamiltonian nor fitting to INS spectra to extract exchange couplings---procedures commonly employed in localized spin model analyses.
The demonstrated agreement between experiment and theory thus provides a firm benchmark for validating our 3D electronic structure models as a realistic description of the actual material.
\\
\section{ACKNOWLEDGMENTS}
\indent This research was supported by the Japan Society for the Promotion of Science through the Grant-in-Aid for Young Scientists 
(Grant Nos.~18K13500 and 19K14666). 
Neutron scattering experiments were performed at the Materials and Life Science Experimental Facility of J-PARC 
under Proposal Nos.~2021I0014 and 2019I0001. NM thanks Dr.\ Tatsuya Kobayashi for carefully reading the manuscript and providing helpful comments.
\bibliography{bibliography}
\end{document}


\title{Supplemental Material for ``Three-dimensional spin susceptibility in Ba$_{0.75}$K$_{0.25}$Fe$_{2}$As$_{2}$: Out-of-plane modulation revealed by neutron spectroscopy and theoretical modeling''}

\author{Naoki Murai}
\affiliation{Materials and Life Science Division, J-PARC Center, Japan Atomic Energy Agency, Tokai, Ibaraki 319-1195, Japan \looseness=-1}
\author{Katsuhiro Suzuki}
\affiliation{Department of Mechanical Engineering, National Institute of Technology, Niihama College, Niihama 792-8580, Japan}
\author{Masamichi~Nakajima}
\affiliation{RIKEN Center for Emergent Matter Science (CEMS), Wako 351-0198, Japan}
\author{Maiko Kofu}
\affiliation{Institute for Solid State Physics, The University of Tokyo, Kashiwa 277-8581, Japan}
\affiliation{Materials and Life Science Division, J-PARC Center, Japan Atomic Energy Agency, Tokai, Ibaraki 319-1195, Japan \looseness=-1}
\author{Seiko~Ohira-Kawamura}
\affiliation{Materials and Life Science Division, J-PARC Center, Japan Atomic Energy Agency, Tokai, Ibaraki 319-1195, Japan \looseness=-1}
\author{Yasuhiro~Inamura}
\affiliation{Materials and Life Science Division, J-PARC Center, Japan Atomic Energy Agency, Tokai, Ibaraki 319-1195, Japan \looseness=-1}
\author{Ryoichi Kajimoto}                                                                           
\affiliation{Materials and Life Science Division, J-PARC Center, Japan Atomic Energy Agency, Tokai, Ibaraki 319-1195, Japan \looseness=-1}

\date{\today}

\maketitle
\section{Neutron diffraction characterization of the co-aligned single crystals}
\begin{figure}[t]
  \begin{center}
    \includegraphics[width=0.90\linewidth, pagebox=artbox]{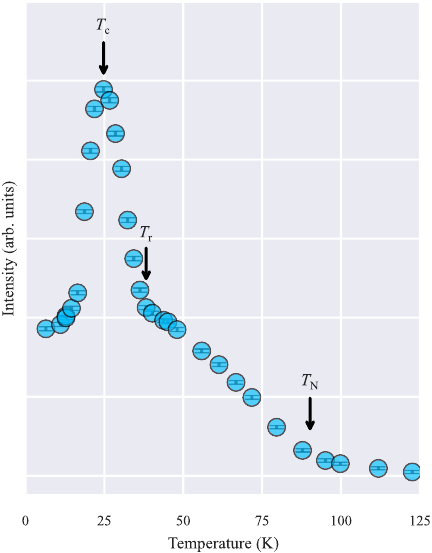}
    \caption{Temperature dependence of the magnetic Bragg intensity at ${\mathbf Q} = (0.5, 0.5, 1)$ for the co-aligned Ba$_{0.75}$K$_{0.25}$Fe$_2$As$_2$ single crystals used in the INS measurements. 
    The magnetic order sets in at $T_\mathrm{N} = 90$~K, followed by a spin reorientation transition at $T_\mathrm{r} = 38$~K. 
    A substantial suppression of the magnetic Bragg intensity is observed below $T_\mathrm{c} = 25$~K, due to the onset of superconductivity and its competition with magnetism.
    }
  \label{sFig1}
  \end{center}
\end{figure}
\indent INS experiments typically require large assemblies of co-aligned single crystals to achieve sufficient signal intensity. 
Consequently, characterizing a randomly selected single crystal using magnetic susceptibility or resistivity measurements does not necessarily represent the properties of the actual sample assembly used in the INS measurements. 
This issue is particularly critical for studies of doped systems, such as the present work, where compositional variations among individual crystals are unavoidable to some extent.
As such, to determine transition temperatures that characterize the entire co-aligned crystal assembly used in the INS experiments, we performed neutron diffraction measurements on the full assembly.
These measurements were carried out on the 4SEASONS TOF chopper spectrometer~\cite{Kajimoto2011, Nakamura2009}, operated in white-beam mode without Fermi chopper operation.
Figure~\ref{sFig1} shows the temperature dependence of the magnetic Bragg intensity at $\mathbf{Q} = (0.5, 0.5, 1)$. 
The onset of magnetic order is observed at approximately $T_{\mathrm{N}} = 90$~K, 
followed by a spin reorientation transition at $T_{\mathrm{r}} = 38$~K, where the magnetic moments switch from in-plane to out-of-plane alignment~\cite{Waber2015}. 
This behavior is consistent with the transition into the so-called $C_{\mathrm{4}}$ magnetic phase, previously reported in Refs.~\cite{Avci2014,Waber2015,Bohmer2015,Taddei2016,Allred2016}. 
Below $T_{\mathrm{c}} = 25$~K, superconductivity emerges, leading to a suppression of the magnetic Bragg intensity due to competition between magnetism and superconductivity.
The clear manifestation of the $C_{\mathrm{4}}$ magnetic phase---which appears only within a narrow doping range ($x \approx 0.25$--$0.28$) in Ba$_{1-x}$K$_x$Fe$_2$As$_2$~\cite{Bohmer2015}---indicates 
a high degree of compositional uniformity across the sample assembly, even though it consists of many individual crystals.
While the sample lies within the $C_{\mathrm{4}}$ magnetic phase, the observed $L$-modulated spin susceptibility---with maxima at odd $L$---is a robust feature shared by both the $C_{\mathrm{4}}$ and $C_{\mathrm{2}}$ phases. 
Minor variations in relative intensity among odd $L$ positions may arise from spin reorientation~\cite{Guo2019}, but these do not alter the central conclusions of the present study.  
Our goal here is to theoretically model the $L$-modulated spin susceptibility, which reflects the out-of-plane AFM instability, based on a realistic 3D electronic structure model.  
Detailed investigations of the spin dynamics unique to the $C_{\mathrm{4}}$ phase will be reported elsewhere.
\begin{figure*}[t]
    \includegraphics[width=0.92\linewidth, pagebox=artbox]{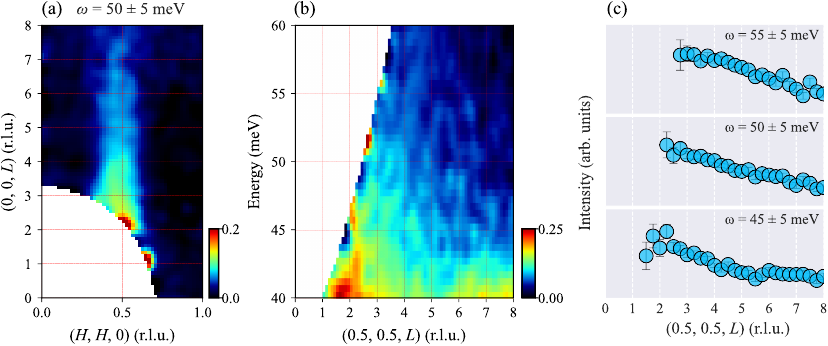}
      \begin{center}
    \caption{Spin excitations measured with $E_i = 125$ meV at $T < 10$ K.
(a) Constant-energy slice in the $(H,H,L)$ plane at $\omega = 50 \pm 5$ meV.
(b) False-color energy spectrum along $(0.5,0.5,L)$.
(c) Representative constant-energy cuts along $(0.5,0.5,L)$ at $\omega = 45 \pm 5$, $50 \pm 5$, and $55 \pm 5$ meV.
In contrast to the pronounced odd-$L$ modulation observed at low energies, the high-energy intensity is essentially $L$-independent, apart from a smooth magnetic form-factor suppression at large $L$, consistent with a predominantly 2D response.
    }
  \label{sFig2}
  \end{center}
\end{figure*}
\section{\boldmath High-energy spin fluctuations measured with $E_i = 125$ meV}
\indent To extend the accessible energy range beyond that covered by $E_i = 55.6$~meV, additional measurements were performed with a higher incident energy of $E_i = 125$~meV.
Constant-energy slices in the $(H,H,L)$ plane at $\omega = 50 \pm 5$~meV are shown in Fig.~\ref{sFig2}(a).
In contrast to the pronounced odd-$L$ modulation observed at low energies, the intensity at this energy becomes nearly uniform along $L$, indicating a strong reduction of the $L$-dependent modulation.
The corresponding energy--$L$ intensity map along $(0.5,0.5,L)$ [Fig.~\ref{sFig2}(b)] shows no discernible odd-$L$ modulation over the measured energy window. 
Instead, the intensity remains essentially uniform along $L$, with the only $L$ dependence being a smooth and monotonic suppression at larger $L$ that follows the expected magnetic form-factor dependence. 
No additional structure beyond the form-factor attenuation is observed.
Representative constant-energy cuts at $\omega = 45$, 50, and 55~meV [Fig.~\ref{sFig2}(c)] further confirm the absence of a pronounced odd-$L$ modulation in this high-energy regime.
These results are consistent with the behavior reported in the main text and support the nearly 2D character of the high-energy spin fluctuations.
\\
\section{Doping dependence of in-plane spin susceptibility}
\begin{figure*}[t]
  \includegraphics[width=1.00\linewidth]{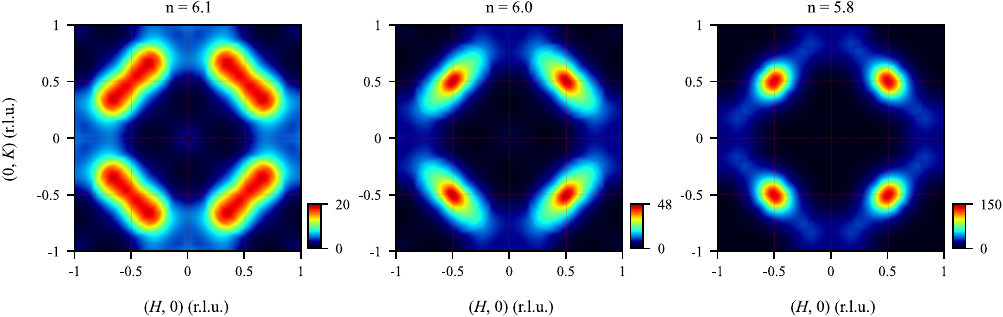}
    \begin{center}
  \caption{
The in-plane spin susceptibility of BaFe$_2$As$_2$ at $\omega = 10$~meV for different electron fillings: $n = 6.1$ (electron-doped), $n = 6.0$ (undoped), and $n = 5.8$ (hole-doped). 
For all fillings, the same set of interaction parameters was used throughout the calculations.
The momentum-space structure of the spin susceptibility evolves with filling, exhibiting a transverse elongation and incommensurate peak splitting on the electron-doped side, whereas the response becomes more isotropic and shifts back toward the commensurate wavevector on the hole-doped side. 
  }
 \label{sFig3}
  \end{center}
\end{figure*}
\indent In FeSCs, the momentum-space structure of the in-plane spin susceptibility is known to evolve systematically with carrier doping~\cite{Suzuki2011PRB,Park2010,Luo2012}.
Since the present study concerns the hole-doped compound Ba$_{0.75}$K$_{0.25}$Fe$_2$As$_2$, it is useful to examine to what extent this doping dependence can be captured within the rigid-band approximation adopted in our calculations. 
To this end, we examined the evolution of the in-plane spin susceptibility within the rigid-band approximation, in which doping is modeled as a rigid shift of the Fermi level.
\\
\indent As shown in Fig.~\ref{sFig3}, on the electron-doped side with a filling of $n = 6.1$, the in-plane spin susceptibility shows transverse elongation and even splits into incommensurate peak positions.
As the filling decreases toward the hole-doped side ($n=5.8$), the susceptibility peak becomes more isotropic and shifts back to the commensurate position at $\mathbf{q}=(0.5,0.5)$. 
The doping-dependent peak shift reflects the relative size mismatch between the hole and electron Fermi surfaces, which evolves in an electron-hole asymmetric manner with doping~\cite{Suzuki2011PRB}. 
By treating the doping effect as a rigid-band shift of the Fermi level, the present calculation qualitatively reproduces the overall doping-dependent trends of the spin susceptibility observed in INS experiments~\cite{Park2010,Luo2012}.
However, its resolution may be insufficient to capture subtle momentum-space features at specific doping levels.
For example, as shown in the main text, the RPA calculation in Fig.~1(d) predicts a slight transverse elongation, whereas the experimental profile in Fig.~1(e) appears nearly isotropic.
Given the sensitivity of the spin susceptibility to the electronic band structure, 
this discrepancy likely reflects subtle deviations from the rigid-band approximation, which omits doping-induced changes in the band structure.
A more precise comparison between experiment and theory would require calculations that go beyond the rigid-band approximation, such as those based on the virtual crystal approximation~\cite{Suzuki2014} or explicit supercell calculations~\cite{Kemper2009}. 
Furthermore, the interaction parameters---taken to be fixed across all fillings---should be chosen in a doping-dependent manner due to their sensitivity to material and doping conditions~\cite{Miyake2010}.
\\
\indent  Despite its simplifications, the rigid-band approach robustly reproduces the instability toward AFM ordering. 
However, the precise anisotropy of the peak profile---whose extreme manifestation corresponds to the onset of incommensurability---is 
highly sensitive to fine details of the electronic structure 
and therefore cannot be quantitatively captured within the present approximation. 
Within these limitations, the present modeling nonetheless captures the essential physics underlying the AFM instability 
and provides a sound basis for interpreting the experimental trends.
\bibliography{bibliography}